\documentclass[twocolumn,showpacs,aps,superscriptaddress,amsmath,amssymb,preprintnumbers]{revtex4}

\usepackage{graphicx}
\usepackage{dcolumn}
\usepackage{bm}
\usepackage{amscd}

\usepackage[T1]{fontenc}

\usepackage{color}
\usepackage{colordvi}

\usepackage{citesort}
\usepackage{xspace}
\usepackage{german}
\newcommand{\RED}[1]{\textcolor{red}{#1}}
\newcommand{\BLUE}[1]{\textcolor{blue}{#1}}

\renewcommand{\Re}{\operatorname{Re}}
\newcommand{\unity}{\ensuremath{{\rm 1 \negthickspace l}{}}}
\newcommand{\Adr}{\operatorname{Ad}}
\newcommand{\adr}{\operatorname{ad}}
\newcommand{\tr}{\operatorname{tr}}
\newcommand{\argmax}{\operatorname{argmax}}
\newcommand{\rank}{\operatorname{rk}}

\newcommand{\ket}[1]{\ensuremath{| #1 \rangle}{}}

\newcommand{\ketbra}[2]{\ensuremath{| #1 \rangle \langle #2 |}{}}

\newcommand{\rep}{\ensuremath{\overset{\rm rep}{=}}}

\newcommand{\grape}{{\sc grape}\xspace}

\begin{document}


\title{Optimal Control for Generating Quantum Gates in Open Dissipative Systems}

\author{T.~Schulte-Herbr{\"u}ggen}\email{tosh@ch.tum.de}
\author{A.~Sp{\"o}rl}
\affiliation{Dept.~Chemistry, Technical University Munich, D-85747 Garching, Germany}

\author{N.~Khaneja}
  \affiliation{Division of Applied Sciences, Harvard University, Cambridge MA02138, USA}

\author{S.J.~Glaser}
\affiliation{Dept.~Chemistry, Technical University Munich, D-85747 Garching, Germany}

\date{\today}


\begin{abstract}
Optimal control methods for implementing quantum modules with least
amount of relaxative loss are devised to give best
approximations to unitary gates under relaxation. 
The potential gain by optimal control using relaxation parameters
against time-optimal control is explored and exemplified in numerical and in algebraic terms:
it is the method of choice to govern quantum systems within
subspaces of weak relaxation whenever the drift Hamiltonian would
otherwise drive the system through fast decaying modes.
In a standard model system generalising decoherence-free subspaces to
more realistic scenarios, open\grape -derived controls realise a {\sc cnot} with fidelities beyond $95$\%
instead of at most $15\%$ for a standard Trotter expansion. As additional benefit it
requires control fields orders of magnitude lower than the bang-bang decouplings in the latter.
\end{abstract}

\maketitle


\section{Introduction}
Using experimentally controllable quantum systems to
perform computational tasks or to simulate other quantum systems
\cite{Fey82, Fey96} is promising: by exploiting quantum coherences,
the complexity of a problem may reduce when changing the setting from
classical to quantum.
Protecting quantum systems against relaxation is therefore
tantamount to using coherent superpositions as a resource.
To this end, decoherence-free subspaces have been applied \cite{ZR97LCW98},
bang-bang controls \cite{VKLabc} have been used for decoupling the system from dissipative
interaction with the environment, while a quantum Zeno approach \cite{MisSud77} may be 
taken to projectively keep the system within the desired subspace \cite{FacPas02}.
Controlling relaxation is both important and demanding~\cite{VioLloyd01,LW03,FTP+05,CHH+06},
also in view of fault-tolerant quantum computing \cite{KBL+01} or dynamic
error correction \cite{KV08}.
Implementing quantum gates or quantum modules experimentally is in fact a challenge:
one has to fight relaxation while simultaneously steering the quantum system
with all its basis states into a linear image of
maximal overlap with the target gate. --- 
Recently, we showed how near time-optimal control by \grape \cite{GRAPE} take pioneering
realisations from their fidelity-limit to the decoherence-limit \cite{PRA07}.

In spectroscopy, optimal control helps to keep the state in slowly relaxing
modes of the Liouville space \cite{KLG,XYOFR04,Poetz06}. In quantum computing, however,
the entire basis has to be transformed. 
For generic relaxation scenarios, 
this precludes simple adaptation to the entire Liouville space: the gain
of going along protected dimensions is outweighed by losses in the orthocomplement.
Yet embedding logical qubits as decoherence-protected subsystem
into a larger Liouville space of the encoding physical system 
raises questions: is the target module
reachable within the protected subspace by admissible controls?

In this category of setting, the extended gradient algorithm open\grape
turns out to be particularly powerful to give best approximations 
to unitary target gates in relaxative quantum systems
thus extending the toolbox of quantum control,
see {e.g.} \cite{Lloyd00, PK02, GZC03, OTR04, GRAPE, PRA05, Tarn05, ST04+06, MSZ+06, dAll08}.
Moreover, building upon a precursor of this work \cite{PRL_decoh},
it has been shown in \cite{PRL_decoh2} that non-Markovian relaxation models can be treated
likewise, provided there is a finite-dimensional embedding such that the embedded system
itself ultimately interacts with the environment in a Markovian way.
Time dependent $\Gamma(t)$ have recently also been 
treated in the Markovian \cite{Rabitz07, Rabitz07b} and non-Markovian regime \cite{Lidar08a}.

Here we study model systems that are
{\em fully controllable} \cite{SJ72JS,BW79,TOSH-Diss,Alt03}, 
i.e.~those in which---neglecting 
relaxation for the moment---to any initial density operator $\rho$,
the entire unitary orbit
$\mathcal{U}(\rho):= \{U\rho\, U^{-1}\, |\, U \; {\rm unitary}\}$
can be reached \cite{AA03} by evolutions under the system Hamiltonian (drift)
and the experimentally admissible controls. Moreover, certain
tasks can be performed within a subspace, e.g. a subspace protected 
totally or partially against relaxation explicitly given in the equation of motion.

\section{Theory}
Unitary modules for quantum computation 
require synthesising
a simultaneous linear image of all the basis states spanning the Hilbert space or subspace
on which the gates shall act. It thus generalises the spectroscopic task
to transfer the state of a system from
a given initial one into maximal overlap with a desired target state.

\subsection{Preliminaries}
The control problem of maximising this overlap
subject to the dynamics being governed by an equation of motion
may be addressed by our algorithm \grape \cite{GRAPE}.
For state-to-state transfer in spectroscopy, one simply refers to
the Hamiltonian equations of motion known as
Schr{\"o}dinger's equation (for pure states of closed systems represented in Hilbert space)
or to Liouville's equation (for density operators in Liouville space) 
\begin{eqnarray}
\dot{\ket{\psi}} &=& -i H\; \ket\psi\\
\dot{\rho} &=& -i\,[H,\rho]\;.
\end{eqnarray}
In quantum computation, however, the above have to be lifted to
the corresponding operator equations, which is facilitated using
the notations $\Adr_U(\cdot):= U(\cdot)U^\dagger$ and $\adr_H\,(\cdot):= [H,(\cdot)]$ 
with $U:=e^{-itH}$ obeying
\begin{equation}\label{eqn:exp_ad}
e^{-it\adr_H}(\cdot) = \Adr_U(\cdot) 
\end{equation}
and using \/`$\circ$\/' for the composition of maps in
\begin{eqnarray}
\dot{U} &=& -i H\; U\\
\tfrac{\rm d}{{\rm d} t}\, {\rm Ad}_U &=& -i \adr_H\;\circ\;\Adr_U\;.
\end{eqnarray}
These operator equations of motion occur in two scenarios
for realising quantum gates or modules $U(T)$ with maximum trace fidelities:
The normalised quality function (setting $N:=2^n$ for an $n$"~qubit system henceforth)
\begin{equation}
f' := \tfrac{1}{N}\, \Re \tr \{U_{\rm target}^\dagger U(T)\} 
\end{equation}
covers the case where overall global phases shall be
respected, whereas if a global phase is immaterial \cite{PRA05} (while the fixed phase relation between
the matrix columns is kept as opposed to ref.~\cite{Tesch04}), 
the quality function
\begin{equation}
f := \tfrac{1}{N}\, \Re \tr \{\Adr_{U_{\rm target}}^\dagger \Adr_{U(T)}^{\phantom{\dagger}}\} = \big| f' \big|^2
\end{equation}
applies. The latter identity is most easily seen \cite{PRA05} in the so-called 
$vec$-representation \cite{HJ2} of $\rho$ where one gets 
the conjugation superoperator $\Adr_{U} = \bar{U}\otimes U$
(with $\bar U$ denoting the complex conjugate) and
the commutator superoperator
$\adr_{H} = \unity\otimes H - H^t\otimes\unity$.

\subsection{Open \grape}
Likewise, under relaxation introduced by the operator $\Gamma$ 
(which may, e.g., take GKS-Lindblad form),
the respective Master equations for state transfer \cite{Alt03}
and its lift for gate synthesis read
\begin{eqnarray}\label{eqn:master}
\dot{\rho} &=& -(i \adr_H \,+\,\Gamma)\; \rho\\
\dot F  &=& -(i \adr_H \,+\,\Gamma)\;\circ\; F \quad.\label{eqn:super_master}
\end{eqnarray}
Again with $N:=2^n$ in $n$"~qubit system,
$F$ denotes a {\em quantum map} in $GL(N^2)$ as linear image over all basis 
states of the Liouville space representing the open system. 
The Lie-semigroup properties of $F(t)$ have recently been elucidated in detail \cite{DHKS08}:
it is important to note that only in the special (and highly unusual) 
case of $[\adr_H \,,\,\Gamma\,]=0$ the map $F(t)$ boils down to a
mere contraction of the unitary conjugation $\Adr_{U}$. 
In general, however, one is faced with an intricate interplay of
the respective coherent ($i\adr_H$) and incoherent ($\Gamma$) part of the time evolution:
it explores a much richer set of quantum maps than contractions of $\Adr_{U}$,
as expressed in \cite{DHKS08} in terms of a $\mathfrak k,\mathfrak p$"~decomposition of the generators
in $\mathfrak{gl}(N^2,\mathbb C)$ of quantum maps. As will be shown below, it is this interplay
that ultimately entails the need for relaxation-optimised control based on the full knowledge
of the Master Eqn.~(\ref{eqn:super_master}), while in the special
case of mere contractions of $\Adr_{U}$, tracking maximum qualities against fixed final times 
(\/`top curves\/', {\em vide infra}, e.g.~Fig.~\ref{fig:compare_all} (a) upper panel) 
obtained for $\Gamma = 0$ plus an estimate on the eigenvalues of $\Gamma$ suffice to 
come up with good guesses of controls.  

\begin{figure}[Ht!]
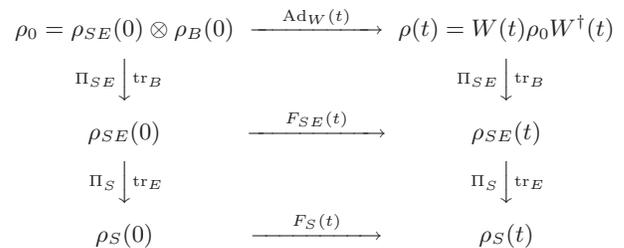

\begin{equation*}
\begin{CD}
{\rho_0 = \rho_{SE}(0)\otimes \rho_B(0)} @>{\quad\Adr_W(t)\quad}>> {\rho(t) = W(t)\rho_0 W^\dagger(t)}\\ 
@V{\Pi_{SE}}V{\tr_B}V              @V{\Pi_{SE}}V{\tr_B}V   \\
{\rho_{SE}(0)}
@>{\quad\;F_{SE}(t)\quad\;}>>
{\rho_{SE}(t)}\\
@V{\Pi_S}V{\tr_E}V              @V{\Pi_S}V{\tr_E}V   \\
{\rho_S(0)} @>{\quad\;\, F_S(t) \quad\;\,}>> {\rho_S(t)}
\end{CD}
\end{equation*}
\caption{\label{fig:CD}
Time-evolution of a quantum system ($S$) embedded in some environment ($E$) and coupled to
a bath ($B$). If the universal system evolves under the global unitary $W(t)$, 
open\grape provides optimal controls for the category of scenarios for which
there is a finite-dimensional embedding such that the embedded system ($SE$) 
follows a time evolution under a {\em Markovian quantum map} $F_{SE}(t)$, 
while the reduced system of concern ($S$) may evolve in a non-Markovian way
by a generic quantum map $F_S(t)$. In the simplest of cases, $\rho_{SE}(0)$ is
of tensor product form with $F_S(t)$ being Markovian itself.
}
\end{figure}
Now for a Markovian Master equation to make sense in terms of physics, it is important that the quantum 
subsystem of concern is itself coupled to its environment in a way justifying to neglect any
memory effects. This means the characteristic time scales under which the environment correlation
functions decay have to be sufficiently smaller than the time scale for the quantum
evolution of the subsystem (see, e.g., \cite{BreuPetr02}). ---
More precisely, as exemplified in Fig.~\ref{fig:CD},
we will assume that either the quantum system itself ($S$) or 
a finite-dimensional embedding of the system ($SE$) can be separated from
the environmental bath ($B$) such that (at least) one of the quantum maps of the reduced system
$F_S(t)$ or $F_{SE}(t)$ is Markovian and allows for a description by a completely
positive semigroup \cite{GKS76,Lind76,Davies76}, if the time evolution for the universal composite of
(embedded) system plus bath is unitary. 
Examples where $F_S(t)$ is Markovian have been given in a precursor \cite{PRL_decoh} to this study,
while a concrete setting of a qubit ($S$) coupled on a non-Markovian scale to a two-level
fluctuator ($E$), which in turn interacts in a Markovian way with a bosonic bath ($B$)
has been described in detail in \cite{PRL_decoh2}.

Henceforth, for describing the method
we will drop the subscript to the quantum map $F(t)$ and tacitly
assume we refer to the smallest embedding such that the map is Markovian
and governed by Eqn.~(\ref{eqn:super_master}). 

Moreover, if the Hamiltonian is composed of the {\em drift term} $H_d$ and
{\em control terms} $H_j$ with piecewise constant {\em control amplitudes} $u_j(t_k)$
for $t_k\in[0,T]$
\begin{equation}
H(t_k) := H_d + \sum_j u_j(t_k) H_j \quad\text{with}\; u_j(t_k)\in \mathcal U \subseteq \mathbb R
\end{equation}
then Eqn.~(\ref{eqn:super_master}) defines a {\em bilinear control system}.

With these stipulations, the \grape algorithm can be lifted to the superoperator
level 
in order to cope with open systems
by numerically optimising
the trace fidelity
\begin{equation}\label{eqn:f_tr}
f_{\rm tr}:= %
  \Re\,\tr\{\Adr_{U_{\rm target}}^\dagger F(T)\}
\end{equation}
for fixed final time $T$. 
For simplicity, we henceforth assume equal time spacing $\Delta t:= t_k-t_{k-1}$ 
for all time slots $k=1,2,\dots,M$,
so $T=M\cdot\Delta t$. Therefore $F(T)=F_M\cdot F_{M-1}\cdots F_k\cdots F_2\cdot F_1$
with every map taking the form
$F_k = \exp\{-(i\adr_{H(t_k)} + \Gamma(t_k))\Delta t\}$
leads to the derivatives
\begin{equation}
\begin{split}
\tfrac{\partial f_{\rm tr}}{\partial u_j(t_k)}& = -\Re\ \tr
\big\{ \Adr^{\dagger}_U \cdot F_M \cdot F_{M-1} \cdots F_{k+1}\;\times\\
        &\times \big(i \adr_{H_j} + \; \tfrac{\partial\Gamma(u_j(t_k))} {\partial u_j(t_k)}\big)
        F_k \Delta t\times F_{k-1} \cdots F_2\cdot F_1 \big\}
\end{split}
\end{equation}
for the recursive gradient scheme
\begin{equation}\label{eqn:recursion}
u_j^{(r+1)}(t_k) = u_j^{(r)}(t_k) + \alpha_r \tfrac{\partial f_{\rm tr}}{\partial u_j(t_k)}\quad,
\end{equation}
where often the uniform $\Delta t$ is absorbed into the step size $\alpha_r >0$.
It gives the update from iteration $r$ to $r+1$ of the control amplitude $u_j$
to control $H_j$ in time slot $t_k$.

\subsubsection*{Numerical Setting}
Numerical open\grape typically started from some $50$ initial conditions to each fixed final
time taking then some $r=10-30 \times 10^3$ iterations (see Eqn.~\ref{eqn:recursion}) to arrive
at one point in the top curve shown as upper trace in Fig.~\ref{fig:compare_all}.

In contrast, for finding time-optimised controls in the closed reference system, 
we used \grape for tracking top curves: this is done by performing optimisations with
fixed final time, which is then successively decreased so as to give a {\em top curve} 
$g(T)$ of quality
against duration of control, a standard procedure used in, e.g., Ref.~\cite{PRA05}.
Finding controls for each fixed final time was typically starting out from some $20$ random initial
control sequences. Convergence to one of the points in time (where Fig.~\ref{fig:compare_all} shows
mean and extremes for a familiy of $15$ different such optimised control sequences) 
required some $r=1000$ recursive iterations each. --- 
Numerical experiments were carried out on single workstations with $512$ MHz to $1.2$ GHz tact rates
and 512 MB RAM.

Clearly, there is no guarantee of finding the global optimum this way, yet the improvements are substantial.

\section{Exploring Applications by Model Systems}
By way of example, the purpose of this section is to demonstrate the power of 
optimal control of open quantum systems as a realistic means for protecting from relaxation. 
In order to compare the results
with idealised scenarios of \/`decoherence-free subspaces\/' and \/`bang-bang decoupling\/', 
we choose two model systems that can partially be tracted by algebraic means.
Comparing numerical results with analytical ones will thus elucidate the pros
of numerical optimal control over previous approaches. --- 
In order to avoid misunderstandings, however, we should emphasize our 
algorithmic approach to controlling open systems (open\grape) is 
{\em by no means limited} to operating within such predesigned subspaces of weak decoherence:
e.g., in Ref.~\cite{PRL_decoh2} we have worked in the full Liouville space of a 
non-Markovian target system.
Yet, not only are subspaces of weak decoherence practically important,
they also lend themselves to demonstrate the advantages of relaxation-optimised
control in the case of Markovian systems with time {\em in}\/dependent relaxation operator~$\Gamma$,
which we focus on in this section.

The starting point is
the usual encoding of one logical qubit in Bell states of two physical ones
\begin{equation}\label{eqn:BellStates}
\begin{split}
{\ket 0}_L &:=\tfrac{1}{\sqrt{2}} \{\ket{01}+\ket{10}\}=\ket{\psi^+}\\
{\ket 1}_L &:=\tfrac{1}{\sqrt{2}} \{\ket{01}-\ket{10}\}=\ket{\psi^-}
\end{split}
\end{equation}
Four elements then span a Hermitian operator subspace protected against $T_2$-type relaxation
\begin{equation}
\mathcal{B}:={\rm span}_{\mathbb R}\,\{\ketbra{\psi^\pm}{\psi^\pm}\}\quad.
\end{equation}
This can readily be seen, since for any $\rho\in\mathcal{B}$
\begin{equation}\label{eqn:zz-protect}
\Gamma_0(\rho): = [zz,[zz,\rho]]=0\quad,
\end{equation}
where henceforth we use the short-hand $zz:=\sigma_z\otimes\sigma_z/2$ and likewise $xx$ as well as
$\unity\mu\nu\unity:=\tfrac{1}{2}\unity_2\otimes\sigma_\mu\otimes\sigma_\nu\otimes\unity_2$
for $\mu,\nu\in\{x,y,z,\unity\}$.
Interpreting Eqn.~\ref{eqn:zz-protect} as perfect protection against $T_2$"~type decoherence is in line
with the slow-tumbling limit of the Bloch-Redfield relaxation by the spin tensor 
$A_{2,(0,0)}:= \tfrac{1}{\sqrt{6}}(\tfrac{3}{2}\,zz  - \mathbf{I}_1 \mathbf{I}_2)$ \cite{EBW87}
\begin{equation}\label{eqn:GammaT2}
\Gamma_{T_2}(\rho): = [A_{2,(0,0)}^\dagger,[A_{2,(0,0)}^{\phantom{\dagger}},\rho]] 
	= \tfrac{9}{24}[zz,[zz,\rho]]=0\;.
\end{equation}

For the sake of being more realistic, the model relaxation
superoperator mimicking dipole-dipole relaxation within the
two spin pairs in the sense of Bloch-Redfield theory 
is extended from covering solely $T_2$-type decoherence to mildly including
$T_1$ dissipation by taking (for each basis state $\rho$) the sum \cite{EBW87}
\vspace{-2mm}
\begin{equation}\label{eqn:GammaT1T2}
\Gamma(\rho)\;:= \sum\limits_{m_1,m_2=-1}^1 
	\big[A_{2,(m_1,m_2)}^\dagger\,,\,\big[A_{2,(m_1,m_2)}^{\phantom{\dagger}}\,,\,\rho\big]\big]\;,
\end{equation}
in which 
the zeroth-order tensor $A_{2,(0,0)}\sim zz$ is 
then scaled 100 times stronger than the new terms. So the resulting 
model relaxation rate constants finally become
$T_2^{-1} : T_1^{-1} = 4.027$ s$^{-1} : 0.024$ s$^{-1} \simeq 170\;:\;1$.

\subsection{Controllability Combined with Protectability against Relaxation}
In practical applications of a given system, a central problem boils down 
to {\em simultaneously} solving two questions: 
(i) is the (sub)system fully controllable and (ii) can the
(sub)system be decoupled from fast relaxing modes while being steered
to the target. 

It is for answering these questions in algebraic
terms that we have chosen the following coupling interactions:
if the two physical qubits are coupled by a Heisenberg"~XX interaction
and the controls take the form of $z$-pulses acting jointly on the two qubits 
with opposite sign, one obtains the usual fully controllable
logical single qubit over $\mathcal{B}$, because
\begin{equation} 
{\langle (z\unity-\unity z), (xx+yy)\rangle}_{\rm Lie}\rep\mathfrak{su}(2)\quad,
\end{equation}
where ${\langle\cdot\rangle}_{\rm Lie}$ denotes the Lie closure under commutation
(which here gives $(yx-xy)$ as third generator to $\mathfrak{su}(2)$).

\subsubsection*{Model System~I}
\begin{figure}[Ht!]
\includegraphics[width=.99\columnwidth]{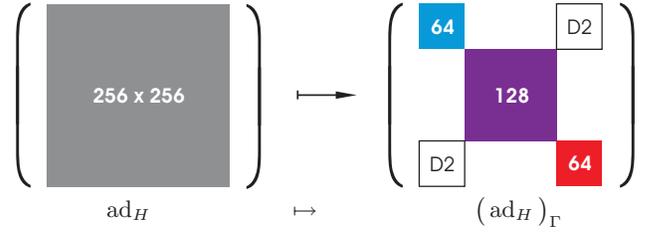}
\mbox{{\phantom{X}\hspace{8mm} $\adr_H\hspace{18mm}\mapsto\hspace{20mm}\big(\adr_H\big{)}_\Gamma$\hspace{8mm}} }
\caption{\label{fig:automorphism} (Colour online)
In a physical four-qubit system for encoding two logical qubits,
the Hamiltonians (in their superoperator representations of $\adr_H$) take the 
form of $256\times 256$ matrices.
In the eigenbasis of $\Gamma$ (Eqn.~\ref{eqn:GammaT1T2}),
the drift Hamiltonian $\adr_{H_{D1}}$ of {\em System~I} block diagonalises 
into slowly relaxing modes (blue) with relaxation rate constants in the
interval $[0\, s^{-1}, 0.060\, s^{-1}]$,
moderately relaxing modes (magenta) with $[4.01\, s^{-1}, 4.06\, s^{-1}]$, 
and fast relaxing modes (red) with $[8.02\, s^{-1}, 8.06\, s^{-1}]$.
In {\em System~II} the
Hamiltonian $\adr_{H_{D1+D2}}$ comprises off-diagonal blocks (empty boxes)
that make the protected modes exchange with the fast decaying ones.
[NB: for pure $T_2$"~relaxation (Eqn.~\ref{eqn:GammaT2}), 
the relaxation-rate eigenvalues would further degenerate to $0\, s^{-1}$ (\/`decoherence-free\/'),
$4\, s^{-1}$ (medium) and $8\, s^{-1}$ (fast) while maintaining the same block structure].\\
}
\end{figure}

By coupling two of the above qubit pairs with an Ising-ZZ
interaction as in Refs.~\cite{LiWu02,WuLi02,ZanLloyd04}
one gets the standard logical two-spin system 
serving as our reference {\em System~I}: it is defined
by the drift Hamiltonian $H_{D1}$ and the control Hamiltonians $H_{C1}, H_{C2}$
\begin{equation}\label{eqn:sys-I}
\begin{split}
H_{D1} :=& J_{xx}\;(xx\unity\unity+\unity\unity xx + yy\unity\unity+\unity\unity yy) + J_{zz}\,\unity zz \unity\qquad\\
H_{C1} :=& z\unity\unity\unity-\unity z \unity\unity\\
H_{C2} :=& \unity\unity z\unity - \unity\unity\unity z\;,\\[-5mm]
\end{split}
\end{equation}
where the coupling constants are set to $J_{xx} = 2$ Hz and $J_{zz} = 1$ Hz.
Hence, over the $T_2$"~decoherence protected subspace spanned by the four-qubit Bell basis 
$\mathcal{B}\otimes\mathcal{B}$ one obtains 
a fully controllable logical two-qubit system
\begin{equation}
{\langle  {H_{D1}}, {H_{C1}}, {H_{C2}}  \rangle}_{\rm Lie}\,\big|_{\mathcal{B}\otimes\mathcal{B}} %
		\rep \mathfrak{su}(4)\quad.
\end{equation}
As illustrated in Fig.~\ref{fig:automorphism}, in the eigenbasis of $\Gamma$ (of Eqn.~\ref{eqn:GammaT1T2})
the Hamiltonian superoperators $\adr_H$ take block diagonal form,
where the first block acts on the Liouville subspace 
$\mathcal{B}\otimes\mathcal{B}$ spanning the states protected against $T_2$-type relaxation.
Thus in more abstract terms (and recalling Eqn.~\ref{eqn:exp_ad}),
the Hamiltonians of System~I restricted to the $T_2$"~protected block,
$\{\adr_{H_{D1}},  \adr_{H_{D1}},  \adr_{H_{D1}}\}\big|_{\mathcal{B}\otimes\mathcal{B}}$,
generate $\Adr_{SU(4)}$ as group of {\em inner automorphisms} over the protected states. 

\subsubsection*{Model System~II}
Now, by extending the Ising-ZZ coupling between the two qubit pairs
to an isotropic Heisenberg-XXX interaction, 
one gets what we define as {\em System~II}. Its drift term
with the coupling constants being set to $J_{xx} = 2$ Hz and $J_{xyz} = 1$ Hz
reads
\begin{equation}\label{eqn:sys-II}
\begin{split}
H_{D1+D2}\;\; :=\;\; &J_{xx}\,\big(xx\unity\unity+\unity\unity xx + yy\unity\unity+\unity\unity yy\big)\\
	     + &J_{xyz}\;\;\,\big(\unity xx \unity + \unity yy \unity + \unity zz \unity\big)
\end{split}
\end{equation}
and it takes the system out of the decoherence-protected subspace
due to the off-diagonal blocks in Fig.~\ref{fig:automorphism}; so the
dynamics finds its Lie closure in a 
much larger algebra isomorphic to $\mathfrak{so}(12)$,
\begin{equation}
\dim {\langle(H_{D1+D2}), H_{C1}, H_{C2} \rangle}_{\rm Lie}  = 66\;,
\end{equation}
to which $\mathfrak{su}(4)$ is but a subalgebra.
\medskip

Note that $e^{-i\pi H_{C\nu}} (H_{D1+D2}) e^{i\pi H_{C\nu}} = H_{D1-D2}$ 
for either $\nu=1,2$.  So invoking Trotter's formula 
\begin{equation}
\lim\limits_{n\to\infty} \big(e^{-i(H_{D1+D2})/(2n)} e^{-i(H_{D1-D2})/(2n)}\big)^n = e^{-i H_{D1}}
\end{equation}
it is easy to see that the dynamics of System~II may reduce to
the subspace of System~I in the limit of infinitely many switchings
of controls $H_{C1}$ or $H_{C2}$ and free evolution under $H_{D1+D2}$.
It is in this {\em decoupling limit} that System~II encodes a fully controllable logical two-qubit
system over the then {\em dynamically protectable basis states} of $\mathcal{B}\otimes\mathcal{B}$.

In the following paragraph we may thus compare the numerical results
of decoherence-protection by optimal control with alternative pulse sequences
derived by paper and pen exploiting the Trotter limit.
As an example we choose the CNOT gate in a logical two-qubit
system encoded in the protected four-qubit physical basis $\mathcal{B}\otimes\mathcal{B}$.

\begin{figure*}[Ht!]
\mbox{\scriptsize{\sf \phantom{X}\hspace{3mm} (a) \hspace{42mm}(b)\hspace{104mm}} }\\[-0mm]
\includegraphics[height= 65 mm]{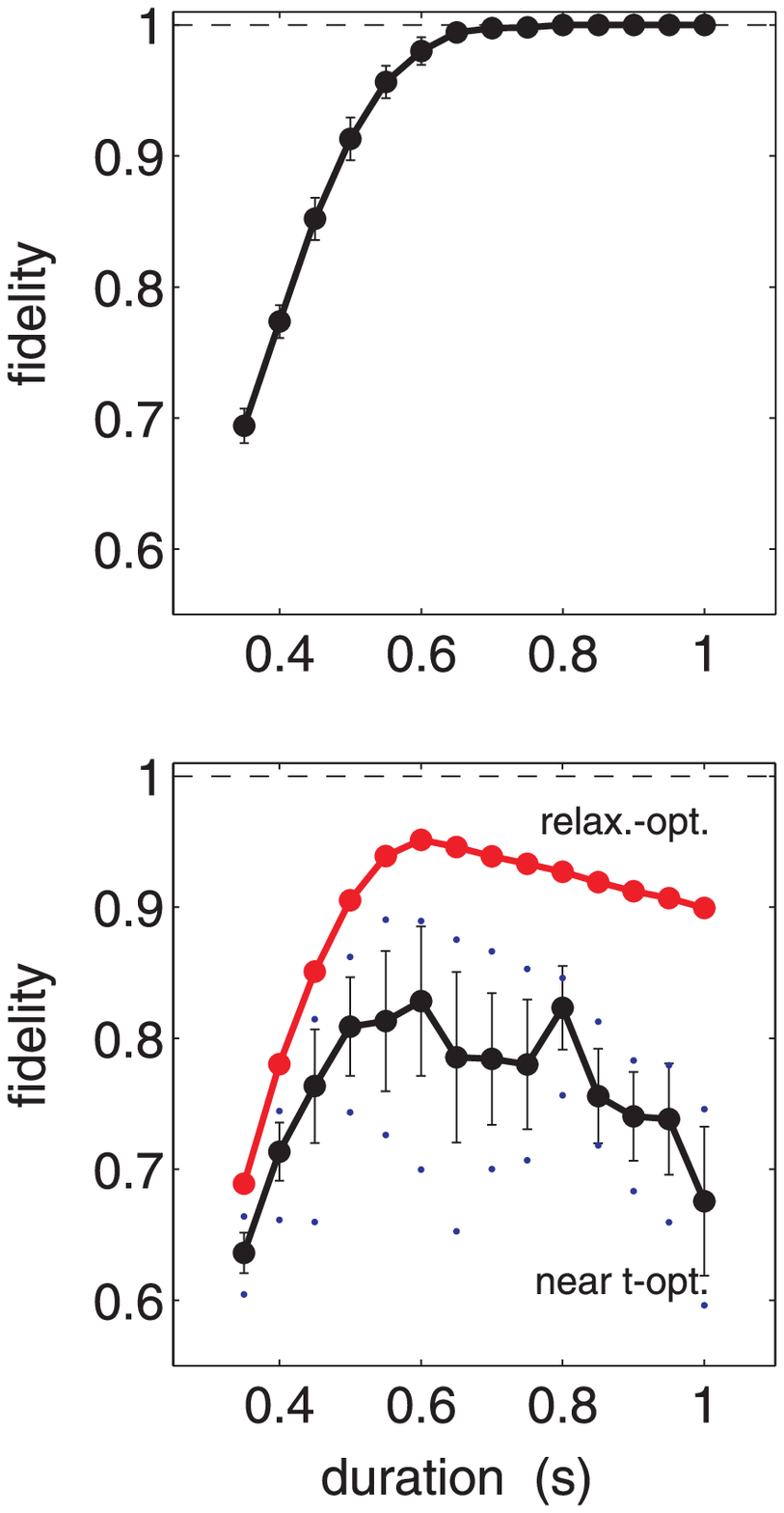}
\hspace{9mm}
\includegraphics[height= 65 mm]{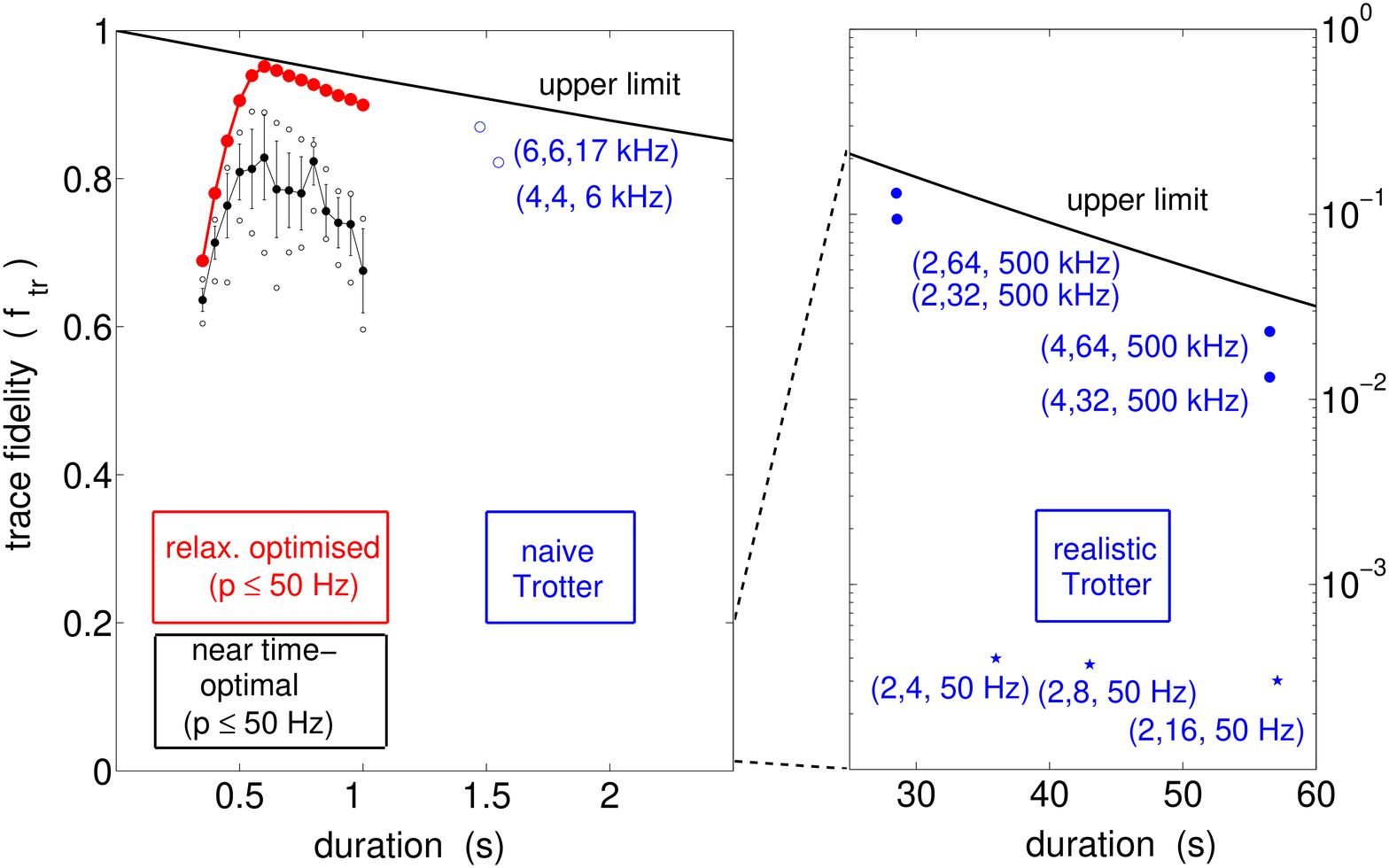}
\caption{\label{fig:compare_all}
(Colour online)
Fidelity of a CNOT gate encoded in an open system of four physical qubits 
in different scenarios of {\em System~II} (see text). 
For reference, the top panel of (a) shows the top curve $g_0(T)$
consisting of maximum obtainable fidelities against fixed final time $T$
in the absence of relaxation; mean and rmsd are shown for families of 15
independent control sequences generated for each $T$. The lower panel of (a)
shows their performance vastly scattering in the presence of relaxation ($\bullet$)
with the intervals giving mean $\pm$ rmsd for all the $15$ control sequences tested
(dots for best and worst values), while numerical optimal control under 
explicit relaxation (\RED{$\bullet$}) is far superior. 
In (b) these results are compared with
(\BLUE{$\mathbf \circ$}) by naive Trotter 
calculations assuming to every interaction the inverse is directly obtainable; 
(\BLUE{$\bullet$}) depicts a realistic Trotter 
approach, where the inverse has to be explicitly generated.
The numbers in brackets $(n_1, n_2, p)$ give the expansion coefficients $n_1, n_2$ of
Fig.~\ref{fig:expansions} and the max.~control power~$p$
(counted as number of $2\pi$"~rotations per second).
Note that numerical optimal control requires some four orders of magnitude less
power than the bang-bang type decoupling from fast relaxing modes used in
the Trotter expansions.
The upper quality limit is imposed by slow $T_1$-type relaxation (see text).
Without relaxation, all the Trotter sequences would
achieve fidelities between $93$ and $99$~\%,
except (\BLUE{$\star$}) the ones limited to control fields of powers $p\leq\,50$ Hz: 
they would fall below $5$\%.
}
\end{figure*}

\subsection{Results on Performing Target Operations under Simultaneous Decoupling}
The model systems are completely parameterised by their respective
Master equations, i.e.~by putting together the Hamiltonian parts of
Eqns.~(\ref{eqn:sys-I}) for System~I or Eqn.~(\ref{eqn:sys-II}) for System~II and the relaxative
part expressed in Eqn.~(\ref{eqn:GammaT1T2}).
We will thus compare different scenarios of approximating the logical
CNOT target gate ($\Adr_{U_{\rm CNOT}}$) by the respective quantum map $F(T)$
while at the same time, the logical two-qubit subsystem has to be decoupled
from the fast decaying modes in order to remain within a weakly relaxing subspace.
This is what makes it a demanding simultaneous optimisation task. --- 
The numerical and analytical results are summerised in Fig.~\ref{fig:compare_all}; 
they come about as follows. 

\subsubsection{Comparison of Relaxation-Optimised and Near Time-Optimal Controls}
\begin{figure}[Hb!]
\mbox{\scriptsize{\sf (a) \hspace{41mm}(b)\hspace{44mm}} }\\[-0mm]
\includegraphics[scale=0.17]{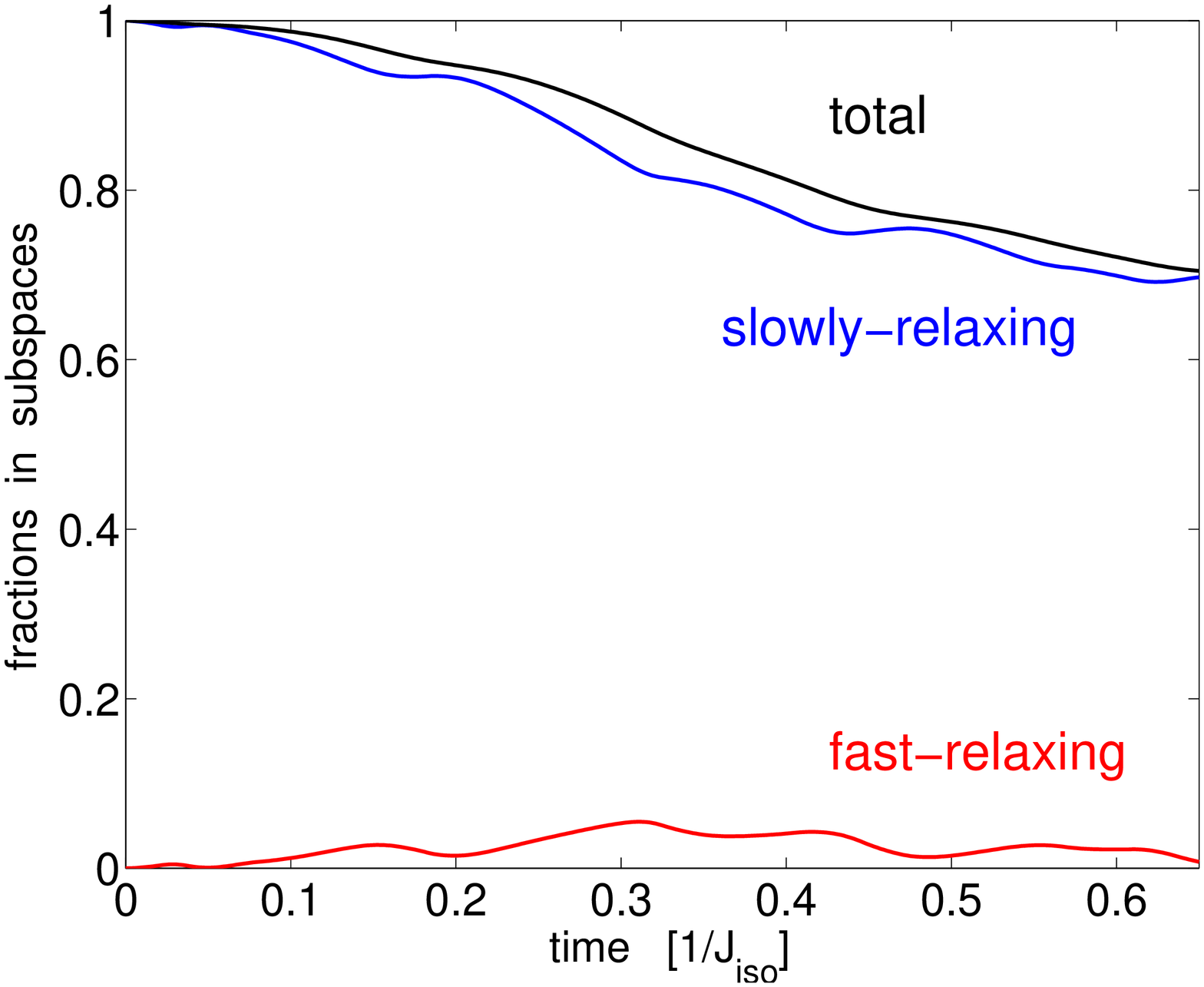}
\includegraphics[scale=0.17]{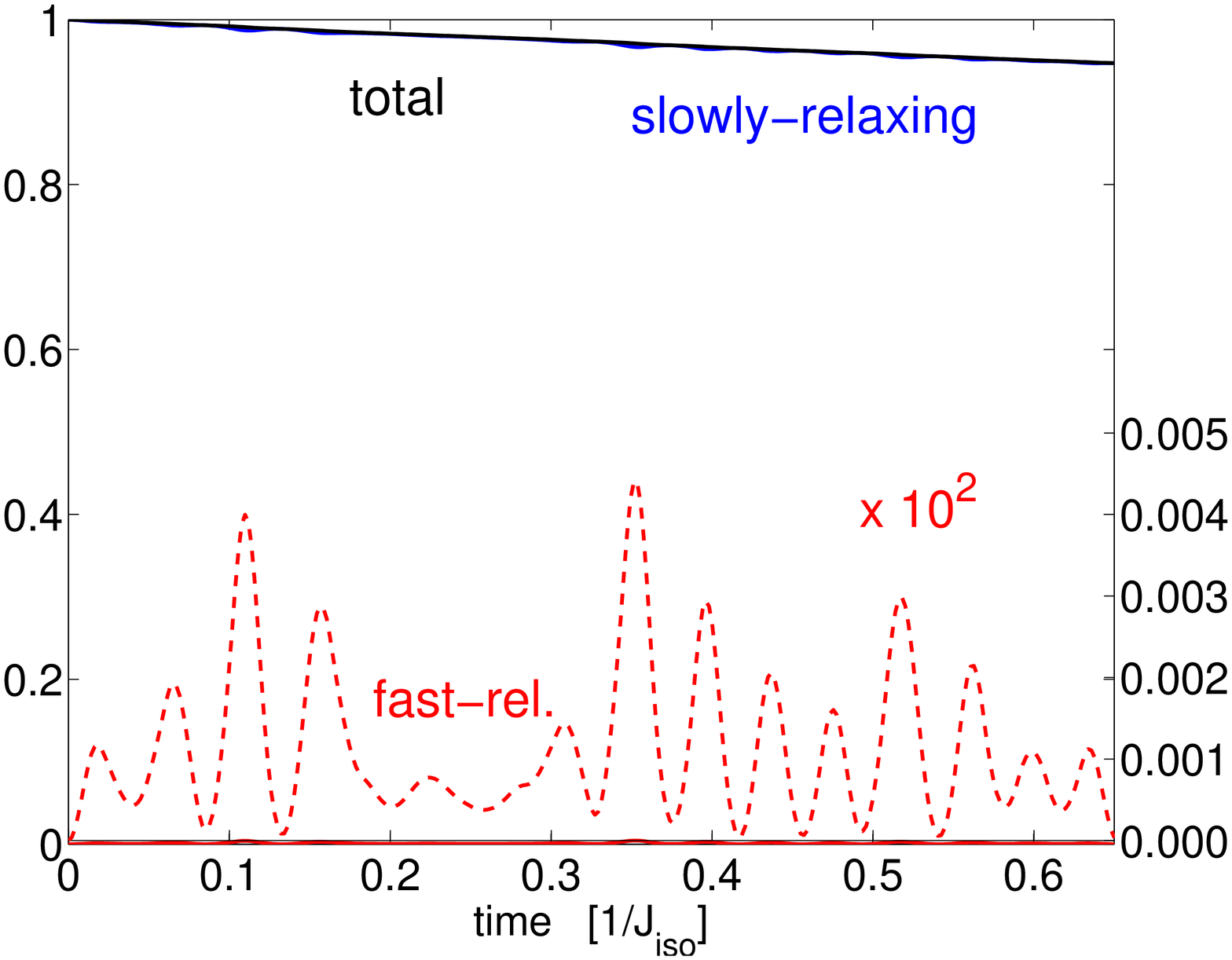}
\caption{\label{fig:compare_t_r}
(Colour online)
(a) Time evolution of all the protected basis states under a typical time-optimised control
of Fig.\ref{fig:compare_all}. Projections into the slowly-relaxing and fast-relaxing
parts of the Liouville space are shown. (b) Same for the new decoherence-avoiding controls.
{\em System~II} (see text) then stays almost entirely within the $T_2$"~protected subspace.
}
\end{figure}

With decoherence-avoiding numerically optimised controls one obtains
a fidelity beyond $95\%$, while near time-optimal controls
show a broad scattering as soon as relaxation is taken into account: 
among the family of 15 sequences generated, serendipity
may help some of them to reach a quality of $85$ to $90\%$, while others perform as bad
as giving $65\%$. With open\grape performing about two
standard deviations better than the mean obtained without taking relaxation into account,
only $2.5 \%$ of near time-optimal control sequences would roughly be expected to reach
a fidelity beyond $95\%$ just by chance.
Fig.~\ref{fig:compare_t_r} then elucidates how the new decoherence avoiding controls
keep the system almost perfectly within the slowly-relaxing subspace,
whereas conventional near time-optimal controls partly sweep through the fast-relaxing subspace
thus leading to inferior quality.

\begin{figure*}[Ht!]
\includegraphics[width=1.7\columnwidth]{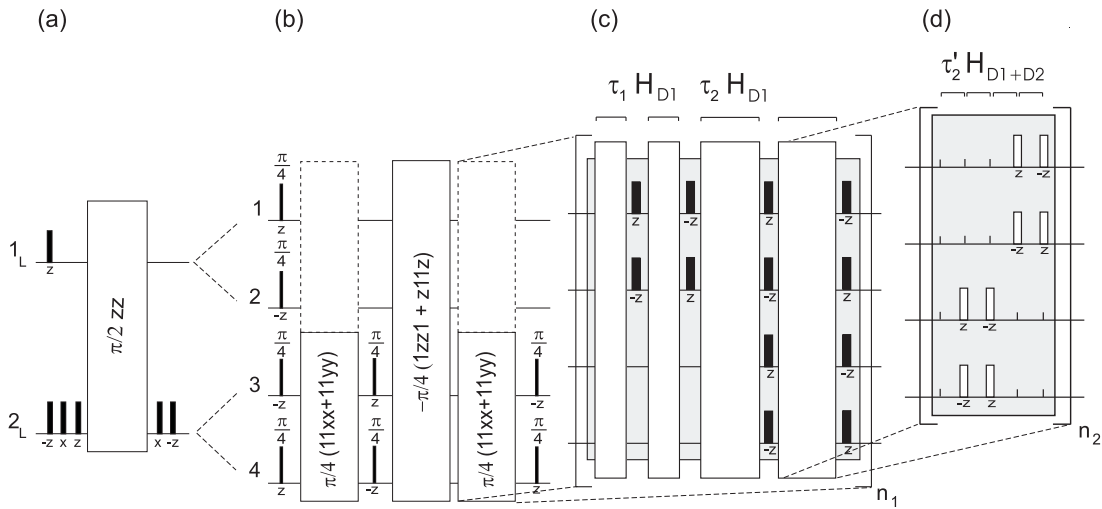}
\caption{\label{fig:expansions}
Impractical academic alternative to numerical control:
the algebraic derivation of controls for a CNOT in the slowly-relaxing subspace 
proceeds from (a) the logical two-qubit system via (b) the encoded schematic
physical four-qubit system to the physical realisations
in the settings of (c) System~I and (d) System~II as defined in the text.
Note that by the definition of the encoding Bell states (Eqn.~\ref{eqn:BellStates})
all angles halve upon going from (a) to (b).
Effective Hamiltonians $\tau H$ are represented by large frames; $n_1, n_2$
indicate repetitions for the respective Trotter expansions.
Black bars are local $\tfrac{\pi}{2}$-pulses with phases given as subscript 
unless other rotation angles given on top; empty bars denote local $z$-rotations by flip angle~$\pi$. 
These pulses can be seen as bang-bang type controls; due to the high repetition rates 
($n_1=2$ but $n_2 = 64$) the Trotter expansions accumulate field strengths of $p\simeq 500$~kHz,
while the pulse shapes from optimal control require $p\leq 50$~Hz.
Expanding the interior Hamiltonian in (b) is even more complicated
(not shown here) 
}
\end{figure*}

\subsubsection{Comparison to Paper-and-Pen Solutions}
Algebraic alternatives to numerical methods of optimal control
exploit Trotter's formula for remaining within the
slowly-relaxing subspace when realising the target, see, e.g. \cite{SVB+05}. 
Though straightforward, they soon become unhandy as shown in
Fig.~\ref{fig:expansions}.
Assuming for the moment that to any evolution under a drift
$H_d$ the inverse evolution under $-H_d$ is directly available,
the corresponding \/``naive\/'' expansions take almost 3 times the length of
the numerical results, yet requiring much stronger control fields 
($1-17$~kHz instead of $50$ Hz)
as shown in Fig.~\ref{fig:compare_all}. 
In practice, however, the inverse is often not immediately reachable,
but will require waiting for periodicity. 
For instance, in the Trotter decomposition of Fig.~\ref{fig:expansions}~(c),
the Ising term $H_{ZZ}:=\unity zz \unity$ as part of the drift Hamiltonian $H_{D1+D2}$
is also needed with negative sign so that all terms governed by $J_{xyz}$ in Eqn.~\ref{eqn:sys-II}
cancel and only the Heisenberg"~XX terms governed by $J_{XX}$ survive.
But $H_{ZZ}$ cannot be sign-reversed directly by the $z$"~controls
in the sense $\unity zz \unity \mapsto -  \unity zz \unity$ since it clearly commutes with
the $z$"~controls. Thus one will have to choose evolution times ($\tau_2$ in Fig.~\ref{fig:expansions})
long enough to exploit
(quasi) periodicity. However, $H_{D1+D2}$ shows eigenvalues 
lacking periodicity within practical ranges altogether. Moreover, the non-zero eigenvalues
of $H_{D1+D2}$ do not even occur in pairs of opposite sign, hence there is no
unitary transform $U: H \mapsto -H = UAU^\dagger$ to reverse them, and {\em a forteriori}
there is no local control that could do so either \cite{PRA_inv}.

Yet, when shifting the coupling to $J_{xx}=2.23$ Hz to introduce a favourable
quasi-periodicity, one obtains almost perfect projection ($f_{\tr}\geq 1-10^{-10}$)
onto the inverse drift evolution of System~II, to wit
$U^{-1}:=e^{+i\tfrac{\pi}{4} H_{D_1+D_2}}$ after $3.98$ sec and onto $-U^{-1}$ after $1.99$ sec.
Thus the identity $\Adr_{(-U^{-1})} = \Adr_{(U^{-1})}$ 
may be exploited
to cut the duration for implementing $\Adr_{U^{-1}}$ to $1.99$ sec.
Yet, even with these facilitations, the total length required
for a realistic Trotter decomposition (with an overall trace fidelity of $f_{\rm tr}\geq 94.1$ \%
in the absence of decoherence) amounts to some $28.5$~sec as shown in Fig.~\ref{fig:compare_all}.
Moreover, as soon as one includes very mild $T_1$-type processes, the relaxation
rate constants in the decoherence-protected subspace are no longer strictly zero
(as for pure $T_2$-type relaxation), but cover the interval $[0\, {\rm s}^{-1}, 0.060\, {\rm s}^{-1}]$.
Under these realistic conditions, a Trotter expansion 
gives no more than $15$\% fidelity, while the new numerical methods allow for realisations beyond $95$\% fidelity
in the same setting (even with the original parameter $J_{xx}=2.0$ Hz).

\section{Discussion}

In order to extract strategies of how to fight relaxation
by means of optimal control, we classify open quantum systems
(i) by their dynamics being Markovian or non-Markovian and 
(ii) by the (Liouville) state space directly representing
logical qubits either directly without encoding or indirectly with
one logical qubit being encoded by several physical ones.
So the subsequent discussion will lead to assigning different
potential gains to different scenarios as summerised in Table~\ref{tab:strategy}.

Before going into them in more detail, recall (from the section on numerical
setting) that the {\em top curve} $g_0(T)$
shall denote the maximum fidelity against final times $T$ as obtained for
the analogous closed quantum system (i.e.~setting $\Gamma=0$) by way of 
numerical optimal control. Moreover, define $T_*$ as the smallest time such
that $g_0(T_*)=1 - \varepsilon$, where $\varepsilon$ denotes some error-correction
threshold. 

First (I), consider the simple case of a Markovian quantum system with no encoding
between logical and physical qubits, and assume $g_0(T)$ has already been determined.
If as a trivial instance (I.a) one had a 
uniform decay rate constant $\gamma$ so $\Gamma=\gamma\unity$,
then the fidelity in the presence of relaxation would simply boil down to
$f(T) = g_0(T)\cdot e^{-T\cdot\gamma}$. 
Define $T'_*:=\argmax\{f(T)\}$ and pick the set of controls
leading to $g_0(T'_*)$ calculated in the absence of relaxation for tracking $g_0(T)$.
In the simplest setting, they would already be \/`optimal\/' without ever having
resorted to optimising an explicitly open system.
More roughly, the time-optimal controls at $T=T_*$ already provide a good approximation
to fighting relaxation if $T_* - T'_*\geq0$ is small, i.e.~if $\gamma >0$ is small.

Next consider a Markovian system without coding, 
where $\Gamma\neq\gamma\unity$ is not fully degenerate (I.b).
Let $\{\gamma_j\}$ denote the set of (the real parts of the) eigenvalues of $\Gamma$. 
Then, by convexity of $\{e^{-\gamma t}\;|\; t, \gamma > 0\}$, the following rough
\footnote{
	Note that for these limits to hold, one has to assume the averaging in the unitary part
	$g_0(T) = \tfrac{1}{N^2} \tr \{\Adr_U^\dagger F_U(T)\}$ and in the dissipative part
	may be performed independently, for which there is no guarantee unless every scalar
	product contributing to $\tr \{\Adr_U^\dagger F_U(T)\}$ is (nearly) equal. 
}
yet useful
limits to the fidelity $f(T)$ obtainable in the open system 
apply
\begin{equation}
g_0(T)\cdot \exp\{- \frac{T}{N^2}\sum_{j=1}^{N^2}\gamma_j\} 
	\lesssim f(T) \lesssim g_0(T)\cdot \frac{1}{N^2} \sum_{j=1}^{N^2} e^{-\gamma_j T}\;.
\end{equation}

Hence the optimisation task in the open system amounts to approximating the
target unitary gate ($\Adr_{U_{\rm target}}$) by the quantum map $F(T)$ resulting from
evolution under the controls subject to the condition that
modes of different decay rate constants $\gamma_j\neq\gamma_k$ are interchanged
to the least possible amount during the entire duration $0\leq t \leq T$ of the controls.
An application of this strategy known in NMR spectroscopy as TROSY \cite{Perv97}
makes use of differential line broadening \cite{Redfield87}
and partial cancellation of relaxative contributions.
Clearly, unless the eigenvalues $\gamma_j$ do not significantly disperse, the advantage
by optimal control under explicit relaxation will be modest, since the potential
gain in this scenario relates to the variance $\sigma^2(\{\gamma_j\})$.

\begin{table}[Ht!]
\begin{center}
\caption{Gain Potential for Relaxation-Optimised Controls {\em versus} 
	Time-Optimised Controls}\label{tab:strategy}
\begin{tabular}{l|cccc}
\hline\hline\\[-1mm]
Category &\phantom{uu}& Markovian &\phantom{u}& non-Markovian \\[2mm]
\hline\\[-1mm]
encoding: 		&&      &&  \\
\; protected subspace \hspace{0mm} && big  && (difficult\footnote{The problem actually roots %
	in finding a viable protected subspace rather than drawing profit from it.}) \\[2mm]
no encoding: 		&&              &&          \\
\; full Liouville space	&& small--medium&&  medium--big \cite{PRL_decoh2} \\[2mm]
\hline\hline
\end{tabular}\hspace{15mm}
\end{center}
\end{table}

The situation becomes significantly more rewarding when moving to the category (II)
of optimisations restricted to a weakly relaxing (physical) subspace 
used to encode logical qubits. A focus of this work has been on showing
that for Markovian systems encoding logical qubits, the knowledge of the 
relaxation parameters translates into significant advantages of relaxation-optimised
controls over time-optimised ones. This is due to a dual effect: open\grape
readily decouples the encoding subsystem from fast relaxing modes while
simultaneously generating a quantum map of (close to) best match to the
target unitary. Clearly, the more the decay of the subspace differs
from its embedding, the larger the advantage of relaxation-optimised control becomes.
Moreover, as soon as the relaxation-rate constants of the protected
subsystem also disperse among themselves, modes
of different decay should again only be interchanged to the least amount necessary---thus
elucidating the very intricate interplay of simultaneous optimisation tasks that
makes them prone for numerical strategies.

In contrast, in the case of entirely unknown relaxation characteristics,
where, e.g.,~model building and system identification of the
relaxative part is precluded or too costly, we have demonstrated that guesses
of time-optimal control sequences as obtained from the analogous
closed system  may---just by chance---cope with relaxation. 
This comes at the cost of making sure a
sufficiently large family of time-optimal controls is ultimately
tested in the actual experiment for selecting among many
such candidates by trial and error---clearly no more than the
second best choice after optimal control under explicitly known relaxation.

In the non-Markovian case, however, it becomes in general very difficult
to find a common weakly relaxing subspace for encoding (II.b): there 
is no Master equation of GKS-Lindblad form, the $\Gamma(t)$ of which could serve
as a guideline to finding protected subspaces. Rather, one would have to
analyse the corresponding non-Markovian Kraus maps for weakly contracted subspaces
allowing for encodings. --- 
However, in non-Markovian scenarios, the pros of relaxation-optimised control already
become significant without encoding as has been demonstrated in \cite{PRL_decoh2}.

\subsection*{Simultaneous Transfer in Spectroscopy}
Finally, note that the presented algorithm also solves (as  a by-product) the problem 
of {\em simultaneous} state-to-state transfer that may be of interest
in coherent spectroscopy \cite{Science98}. 
While Eqns.~\ref{eqn:super_master} and \ref{eqn:f_tr} refer to the full-rank
linear image $F$, one may readily project onto the states of concern by the appropriate
projector $\Pi$ to obtain the respective dynamics and quality factor of the subsystem
\begin{eqnarray}
\Pi\;\dot{F} &=& - \Pi\;(i \adr_H \,+\,\Gamma)\;\circ\; F\\
f^{(\Pi)}_{\rm tr} &=&  \tfrac{1}{\rank \Pi}\;\Re\,\tr\{\Pi^t\;F_{\rm target}^\dagger \Pi\;F(T)^{\phantom{\dagger}}\}
\end{eqnarray}
reproducing Eqn.~\ref{eqn:master} in the limit of $\Pi$ being a rank"~1 projector.
While such rank"~1 problems under relaxation were treated in \cite{KLG}, the
algorithmic setting of open\grape put forward here allows for projectors of arbitrary rank,
e.g., 
$1\leq \rank \Pi \leq N$ for $n$ spin"~$\tfrac{1}{2}$ qubits with $N:=2^n$.
Clearly, the rank equals the number of orthogonal
state-to-state optimisation problems to be solved {\em simultaneously}.

\section{Conclusions and Outlook}
We have provided numerical optimal-control tools
to systematically find near optimal approximations to  unitary target modules
in open quantum systems.
The pros of relaxation-optimised controls over time-optimised ones depend
on the specific experimental scenario. We have extensively discussed strategies
for fighting relaxation in Markovian and non-Markovian settings with and without
encoding logical qubits in protected subspaces. Numerical results have been
complemented by algebraic analysis of controllability in protected subspaces
under simultaneous decoupling from fast relaxing modes.

To complement the account on non-Markovian systems in \cite{PRL_decoh2},
the progress is quantitatively exemplified 
in a typical Markovian model system of four physical qubits encoding two logical ones:
when the Master equation is known, the new method is systematic and significantly
superior to near time-optimal realisations, which in turn are but a
guess when the relaxation process cannot be quantitatively
characterised. In this case, testing a set of $10-20$ such near time-optimal control
sequences empirically is 
required for getting acceptable results with more confidence, 
yet on the basis of trial and error.
As follows by controllability analysis, Trotter-type expansions allow for
realisations within slowly-relaxing subspaces in the limit of infinitely
many switchings.
However, in realistic settings for obtaining inverse interactions,
they become so lengthy that they only work in the idealised limit of both
$T_2$ and $T_1$-decoherence-free subspaces, but fail
as soon as very mild $T_1$-relaxation processes occur.

Optimal control tools like open\grape are therefore the method of choice
in systems with known relaxation parameters. They accomplish 
decoupling from fast relaxing modes with several orders
of magnitude less decoupling power than by typical bang-bang controls. Being
applicable to spin and pseudo-spin systems, they
are anticipated to find broad use for fighting relaxation in practical quantum
control. In a wide range of settings 
the benefit is most prominent when encoding the logical system in a
protected subspace of a larger physical system.
However, the situation changes upon shifting to a timevarying
$\Gamma(t)$ \cite{Rabitz07}, 
or to more advanced non-Markovian models with $\Gamma\big(u(t)\big)$
depending on time via the control amplitudes $u(t)$ on timescales
comparable to the quantum dynamic process.
Then the pros of optimal control extend to the entire Liouville space,
as shown in \cite{PRL_decoh2}.

In order to fully exploit the power of optimal control of open systems
the challenge is shifted to (i) thoroughly understanding the relaxation mechanisms
pertinent to a concrete quantum hardware architecture and (ii) being able to determine its
relaxation parameters to sufficient accuracy.

\begin{acknowledgments}
This work was
presented in part at the conference {\sc pracqsys}, Harvard, Aug.~2006.
It was supported by the integrated {\sc eu} project {\sc qap} as well as by 
{\em Deutsche Forschungsgemeinschaft}, {\sc dfg}, in {\sc sfb}~631.
Fruitful comments on the e-print version by the respective groups of
F.~Wilhelm, J.~Emerson and R.~Laflamme during a stay at {\sc iqc}, Waterloo as well as by
B.~Whaley on a visit to {\sc uclb} are gratefully acknowledged.
\end{acknowledgments}
\bibliography{control21}
\end{document}